\documentclass[conference]{IEEEtran}
\IEEEoverridecommandlockouts
\usepackage{cite}
\usepackage{amsmath,amssymb,amsfonts}
\usepackage{float}
\usepackage{subfig}
\usepackage{ulem}
\usepackage{graphicx}
\usepackage{textcomp}
\usepackage{xcolor}
\usepackage[pagewise,switch]{lineno}
\usepackage{booktabs}
\usepackage{multirow}
\usepackage{times}
\usepackage[ruled,linesnumbered,vlined]{algorithm2e}
\usepackage{indentfirst}
\usepackage{amsmath}
\usepackage{float}
\usepackage{soul}
\usepackage{color}
\usepackage{caption}
\usepackage{amsthm}
\usepackage{enumitem}
\usepackage{url}

\begin{document}

\title{WiFi-based Real-time Breathing and Heart Rate Monitoring during Sleep}

\author{
\IEEEauthorblockN{Yu Gu\IEEEauthorrefmark{1}, Xiang Zhang\IEEEauthorrefmark{1}, Zhi Liu\IEEEauthorrefmark{4} and  Fuji Ren\IEEEauthorrefmark{2} \\
\IEEEauthorblockA{\IEEEauthorrefmark{1} School of Computer and Information, Hefei University of Technology, China\\
\IEEEauthorblockA{\IEEEauthorrefmark{4} Dept. of Mathematical and Systems Engineering, Shizuoka University, Japan \\}
\IEEEauthorblockA{\IEEEauthorrefmark{2} Dept. of Information Science \& Intelligent Systems, Univ. of Tokushima, Japan\\}
}}}

\maketitle
\begin{abstract}

Good quality sleep is essential for good health and sleep monitoring becomes a vital research topic.
This paper provides a low cost, continuous and contactless WiFi-based vital signs (breathing and heart rate) monitoring method.
In particular, we set up the antennas based on Fresnel diffraction
model and signal propagation theory, which enhances the detection of weak breathing/heartbeat motion.
We implement a prototype system using the off-shelf devices and a real-time processing system to monitor vital signs in real time.
The experimental results indicate the accurate breathing rate and heart rate detection performance.
To the best of our knowledge, this is the first work to use a pair of WiFi devices and omnidirectional antennas to achieve real-time individual breathing rate and heart rate monitoring in different
sleeping postures.

\end{abstract}
\begin{IEEEkeywords} Wi-Fi, CSI, sleep monitoring \end{IEEEkeywords}

\section{Introduction}
\label{Sect:int}
Good quality sleep is essential for good health and well-being\cite{world2004technical}.  However, many diseases in sleep are fatal such as cardiac arrest\cite{pedersen2004breathing}, sleep apnea\cite{min2010noncontact}, and asthma\cite{braun2012bridging}. A recent study also shows that disordered breathing is a significant cause of Sudden Infant Death Syndrome (SIDS) for sleeping infants\cite{facco2014sleep}. In many instances, patients with the respiratory disease only show the symptoms for a short period or at casual occasions, and long-term hospitalization is undoubtedly unrealistic for patients with these diseases. Hence, continuous and cost-effective vital signs (i.e.\textcolor{blue}{,} breathing and heart rate) monitoring at home environment is essential.

Most of the traditional solutions use sensors for physiological signal detection, for example, Polysomnography \cite{kushida2005practice} and Electrocardiogram \cite{he2018litenet}. However, these programs are not suitable for the home environment. Solutions based on pressure or acceleration sensors need to be in contact with the body, and lighting conditions limit computer vision-based solutions. Recently, Radio Frequency (RF) based monitoring solutions \cite{salmi2011propagation,adib2014multi} have drawn considerable attention as they provide non-invasive breathing rate monitoring. However, the equipment used in these solutions is usually expensive and thus can not be widely used.

WiFi-based solutions are recently emerging as an alternative. For example, Liu et al.\cite{liu2016contactless} extract the personal respiratory data by use STFT(Short Time Fourier Transform) on the CSI (Channel State Information) derived from the wireless network card. To obtain a person's breathing information in different sleeping postures, they need to deploy two routers and three computers. \cite{liu2018monitoring} uses a pair of devices to monitor the breathing rate in different sleeping postures. They can get heart rate under strict restrictions, the line of sight between the WiFi device/AP must crossing the person’s chest and use directional antenna). However, they use the metronome to control the person's breathing during the experimental evaluation, instead of letting the tester breathe normally. \cite{wang2016human} and \cite{zhang2018fresnel} calculate the best position for detecting the respiration by deriving the Fresnel diffraction model. However, this is an ideal derivation, while Fresnel theory is limited by other factors like receive antennas and obstacles in the real environment. The limitations of these works and the differences between our work are illustrated in Table \ref{compare}.

Some WiFi-based breath detects methods use Fresnel zone model to help to design the system. For example, \cite{wang2016human} combines the Fresnel model and WiFi radio propagation theory to investigate the impact of human respiration on the receiving CSI. They develop the theory to link one's breathing depth, location, and orientation to the detectability of respiration. Their research finds that CSI performs well in the middle of a Fresnel zone, but terribly at the boundaries. They also observe that the closer a subject is to the Tx/Rx, the better CSI can improve performance. When the direction of the human body is 0 degree, the detection efficiency is the best, and as the angle increases, the effective displacement becomes smaller, and the detection effect gets worse.
\cite{zhang2018fresnel} leverages Fresnel model to researched the relation of body position and the first Fresnel zone, they have detailed formula derivation and experiments to verify the point that: the closer a subject is to the Tx/Rx, the stronger CSI signals can obtain when in the sleep state. These two papers are beneficial for studying the use of WiFi to collect breathing or other micro-motion information; however, the detailed theoretical guidance to the experimental setup is not provided.

\begin{table*}
	\centering   	
	\caption{\label{compare} \upshape The latest research work compares with our system}
	\begin{tabular}{|c|p{3cm}|c|p{4.5cm}|c|c|}
		\hline
		Reference & Vital signs & Real-time & Performance & Equipment &Theory Support\\
		\hline
		\cite{liu2016contactless} & breathing rate(various sleep postures & no & greater than 85\% & 2 transmitters and 3 receivers & no\\
		\hline
		\cite{liu2018monitoring} & breathing rate(various sleep postures) and heart rate (only supine)& no & 80\% estimation errors are less than 0.5bpm for breath rate, 90\% of estimation errors are less than 4bpm for heart rate & pair of transceivers & no\\
		\hline
		\cite{wang2016human} &  breathing rate(various sleep postures)& no & Na & pair of transceivers & yes\\
		\hline
		\cite{zhang2018fresnel} &  breathing rate(various sleep postures)& no & For the good positions, the overall estimation accuracy is as high as 98.8\%. For bad positions, the accuracy decreases to 61.5\%.& pair of transceivers & yes\\
		\hline
		Our System &  breathing rate and heart rate (various sleep postures)& yes & 96.636\% for breathing rate and 94.215\% for heart rate& pair of transceivers & yes\\
		\hline
	\end{tabular}
\end{table*}

This paper aims to provides a low cost, continuous and contactless WiFi based vital signs (i.e. breathing and heart rate) monitoring method, based on our previous experiences \cite{Gu2018EmoSense, Gu17IoT}. In particular, human breathing and heartbeat can result in weak motions in the abdomen and chest. These motions can have some effect on the propagation of WiFi signals and the WiFi CSI can record these effects. We set up the systems to enhance these effects based on Fresnel diffraction models and signal propagation theory, and extract CSI from WiFi physical layer to obtain vital signs.
The main contributions of our work are summarized as follows:
\begin{enumerate}
    \item We are the first to use a pair of WiFi devices and omnidirectional antennas to achieve real-time individual breathing rate and heart rate monitoring in different sleeping postures.
    \item We set up the antennas based on Fresnel diffraction model and signal propagation theory, which enhance the detection of weak breathing/heartbeat motions.
    \item We implement a prototype system and a real-time processing system to monitor vital signs in real-time. The experimental results indicate the proposed scheme's accurate detection performance.
\end{enumerate}

We organize the remainder of this paper as follows: we provide an overview of our preliminary work in Section \ref{Sect:pre}, and introduce the system design in Section \ref{Sect:sys}. Then, we show our experimental details and the experimental results in Section \ref{Sect:per}. Finally, we conclude our work and discuss the open issues in Section \ref{Sect:con}.

\section{Preliminaries}
\label{Sect:pre}

\subsection{Fresnel Zone}
\begin{figure}
	\centering
	\includegraphics[width=0.9\columnwidth]{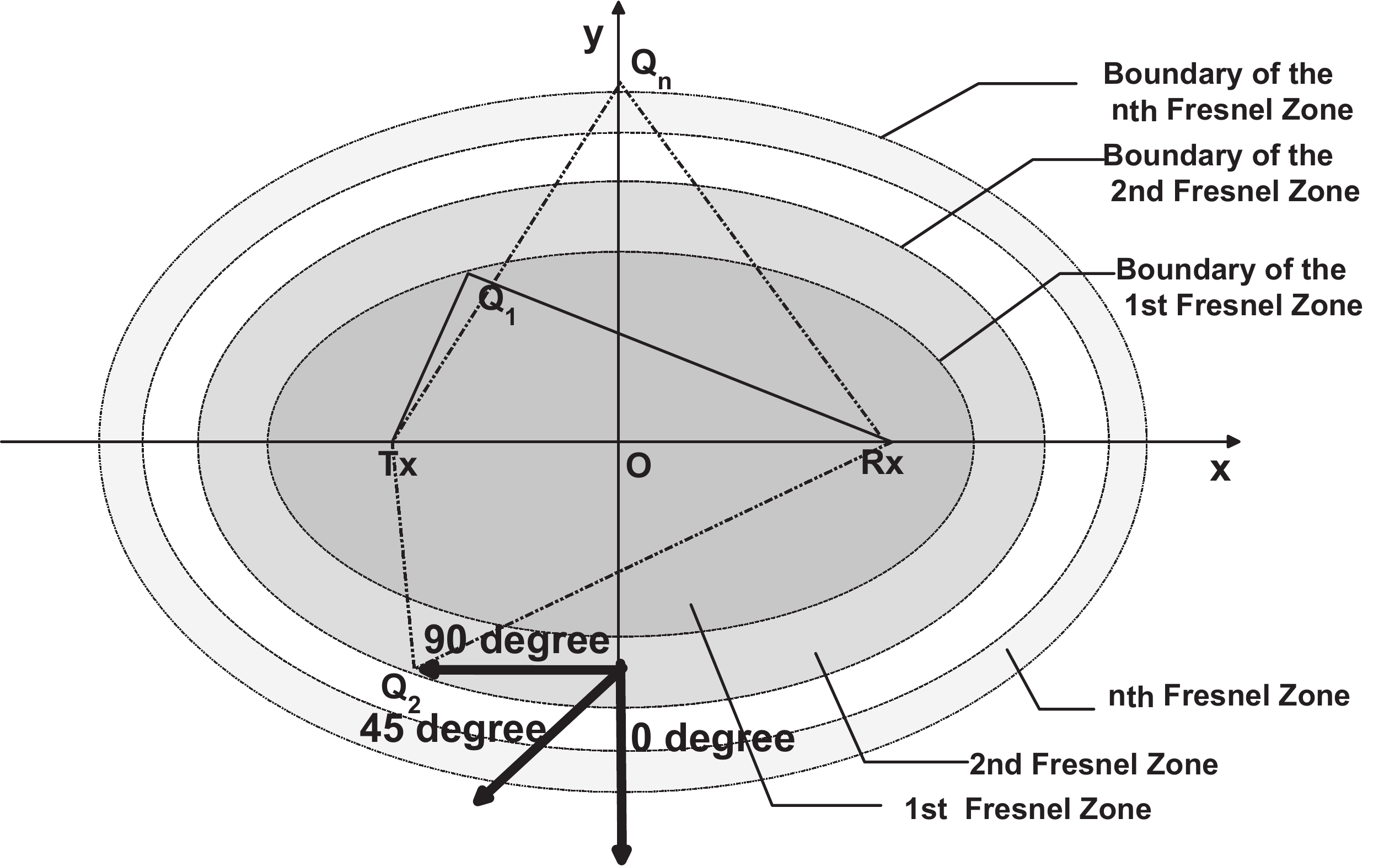}
	\caption{Fresnel Zone}
	\label{fig:fz}
\end{figure}

As shown in Fig. \ref{fig:fz}, Fresnel zones are defined as a series of concentric ellipsoids, and $P_1$ and $P_2$ are the positions of the transmitting antenna and receiving antenna, respectively. $Tx$ and $Rx$ represent the sender and receiver, respectively. For a given radio wavelength $\lambda$, we could construct Fresnel zones by the following equation:
\begin{equation}
\label{equ:fz}
|TxQ_{n}|+|Q_{n}Rx|-|TxRx|=n\lambda/2
\end{equation}
where $Q_n$ is a point at the boundary of the $n$th Fresnel zone.

CFR (channel frequency response) can be expressed simply as the superposition of dynamic path CFR and static CFR, and it can be represented as:
\begin{equation}
\label{equ:CFRSUM}
H(f,t)=H_s(f,t)+H_d(f,t),
\end{equation}

The dynamic CFR can be written as:
\begin{equation}
\label{equ:CFR}
H_d(f,t)=\sum_{k\in D} h_k(f,t) e^{-j2\pi f\tau _k(t)},
\end{equation}
where $f$ and $\tau _k(t)$ represent the carrier frequency and the propagation delay on the $k^{th}$ path, respectively. $D$ is the set of dynamic paths. A subject can reflect a WiFi signal, and if the subject moves a small distance, it leads to changes in the phase of the WiFi signal on the corresponding path. If the subject moves $d(t)$, since wireless signals travel at the speed of light, denoted as $c$, $\tau _k(t)$ can be represented as $d(t)/c$. Let $\lambda$ represent the wavelength, where $\lambda=f/c$. Thus, the phase shift can be written as $e^{-j2\pi d(t)/\lambda}$.

When a subject appears at the boundary of the even/odd Fresnel zone, the dynamic path phase shift $\Delta p$ is equal to $ \pi$ and $2\pi$, respectively. As a result, the combined signal amplitude should be \textbf{degraded} in the even zones and \textbf{enhanced} in the odd zones. In other words, if an object continues to pass through multiple Fresnel zones, the amplitude of the received CSI signal exhibits a sine-like wave.

The larger the effective displacement of the action, the more obvious the response is on the CSI waveform. The effective displacement is along the direction of the normal line, which can cause the reflection path length change.

\subsection{Preliminary Experiments}
The trunk movements caused by breathing and heartbeat are tiny. To explore the possibility of using WiFi signals to detect vital signs, we build a prototype system to carry out some preliminary experiments.

\noindent \textbf{[Prototype]} Our prototype system consists of two commodities MiniPCs, and they are all equipped with an Intel Network Interface Controller 5300. They are respectively as the transmitting and receiving device. Antennas setting shown in Fig.2.

\noindent \textbf{[Participant]} One 22 years old male student participates in our preliminaries experiments.

\noindent \textbf{[Environment]} We conduct the experiments in a  $7\times 10 m^2$  office room, with the furniture including chairs, couches, computer desks, and book cabinets. Other students are also in the same place during the experiments.

\noindent \textbf{[Setting]} The package sending rate is set to be 500hz, participant conducts the experiments with different sleeping postures (Lying down, lying on the side and lying with face down) with varying settings of the antennas (as shown in Fig.2).

\begin{figure}
	\centering
	\subfloat[]{\label{setting1}
		\begin{minipage}{0.43\linewidth}
			\centering
			\includegraphics[width=1\textwidth]{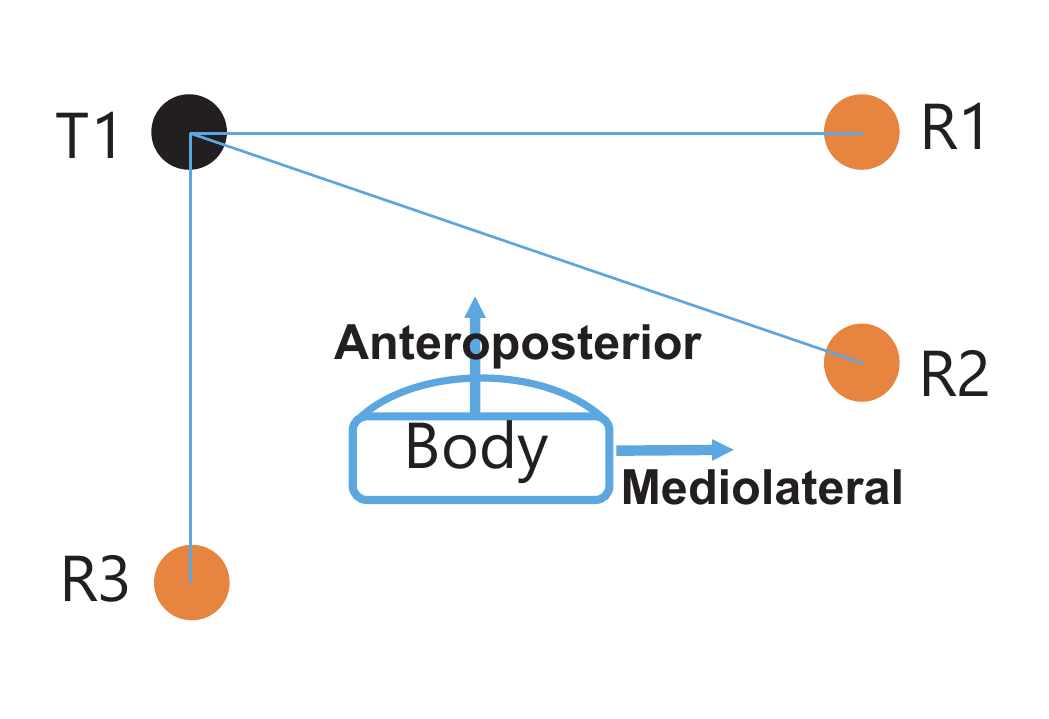}
		\end{minipage}
	}	
	\quad
	\subfloat[]{\label{setting2}
		\begin{minipage}{0.43\linewidth}
			\centering
			\includegraphics[width=1\textwidth]{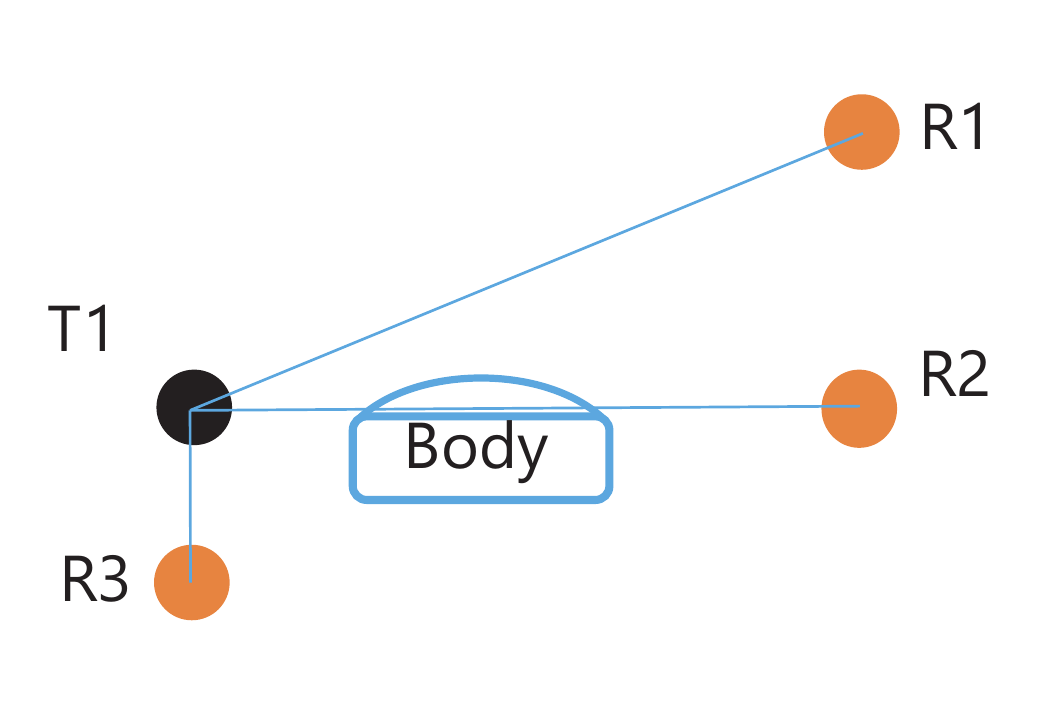}
		\end{minipage}
	}
	
	\quad
		\subfloat[]{\label{setting3}
		\begin{minipage}{0.43\linewidth}
			\centering
			\includegraphics[width=1\textwidth]{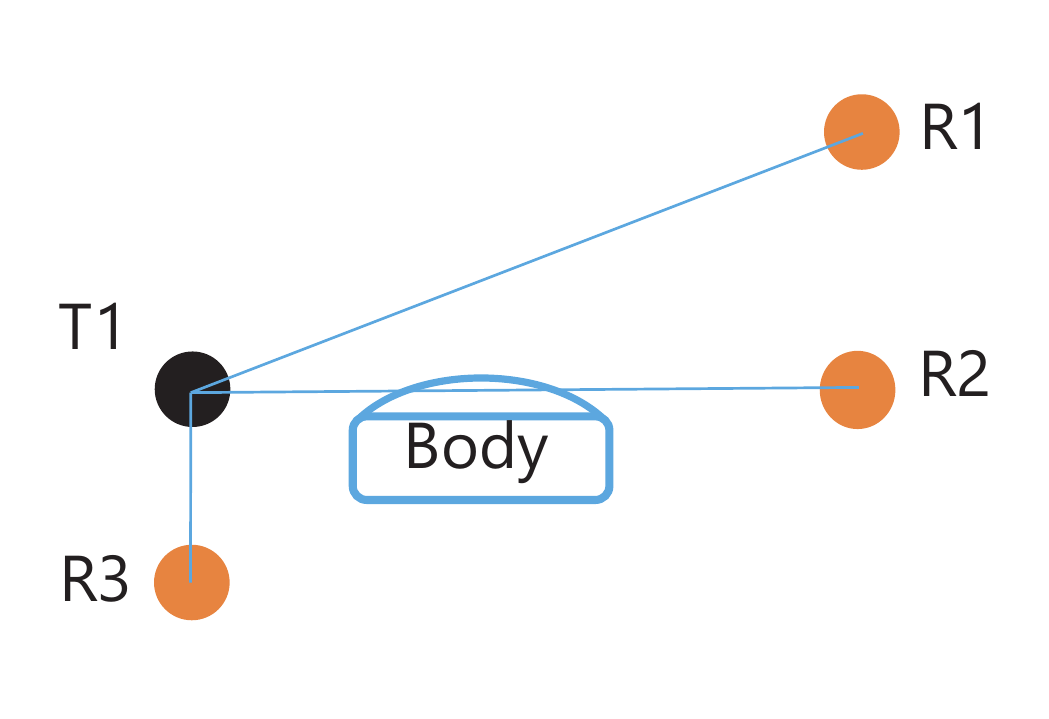}
		\end{minipage}
	}	
	\quad
    \caption{(a) Setting1; T1 is transmit antenna and R1,R2,R3 are receive antennas, the distance from T1 to R3 is 80cm, the distance between T1 and R1 is 120cm; (b) Setting2; the chest is on the LOS of T1-R2, the distance from T1 to R3 is 20cm, and the distance between T1-R2 is 120cm; (c) Setting3; the chest is in the first Fresnel zone of T1-R2, the distance from T1  to R3 is 30cm, and the distance between T1 and R2 is 120cm.}
\end{figure}

We analyze the preliminary results and obtain the following key observations:

\begin{figure*}
	\centering
	\subfloat[]{\label{plan1}
		\begin{minipage}{0.47\linewidth}
			\centering
			\includegraphics[width=1\textwidth]{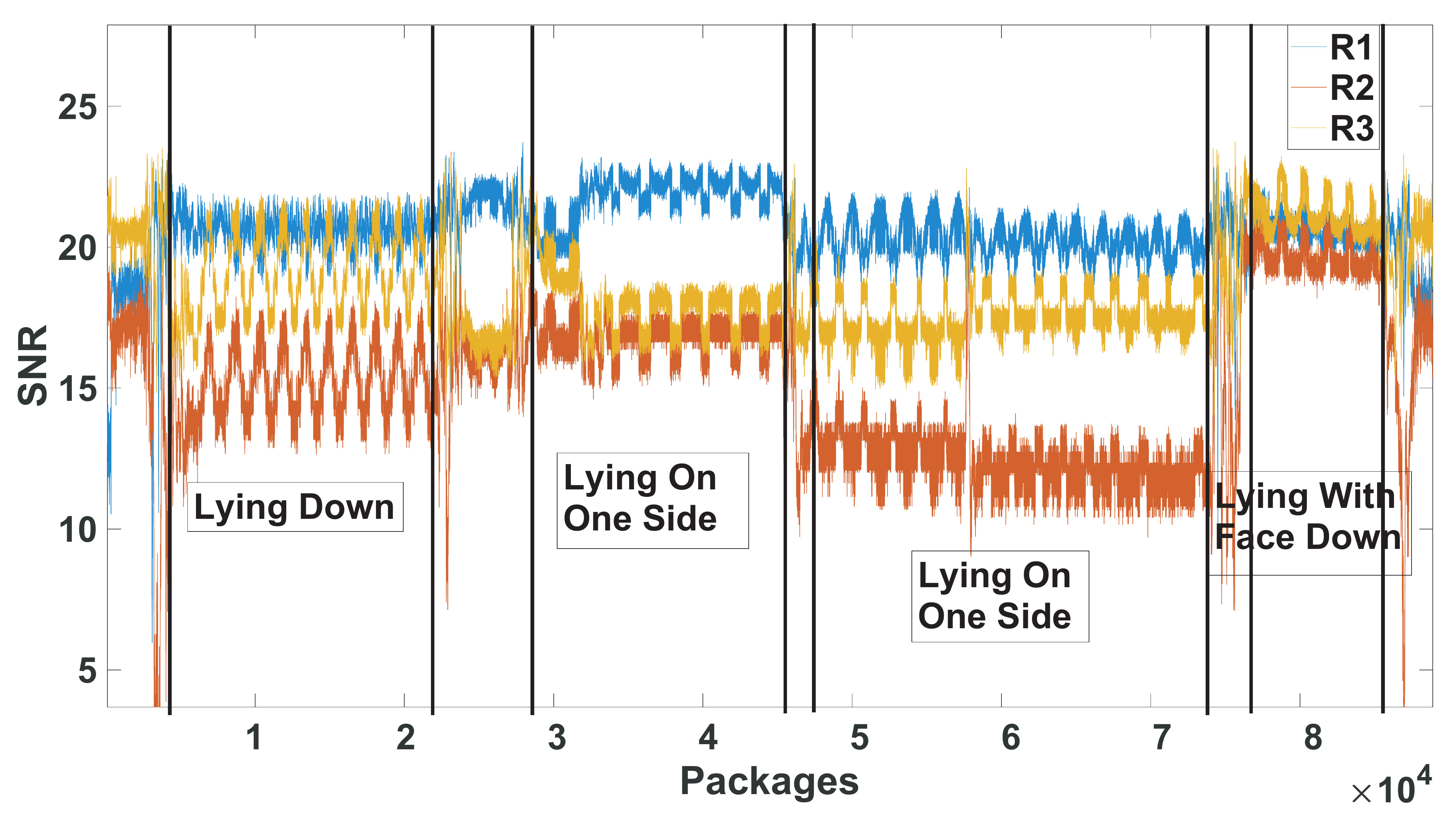}
		\end{minipage}
	}	
	\quad
	\subfloat[]{\label{plan2}
		\begin{minipage}{0.47\linewidth}
			\centering
			\includegraphics[width=1\textwidth]{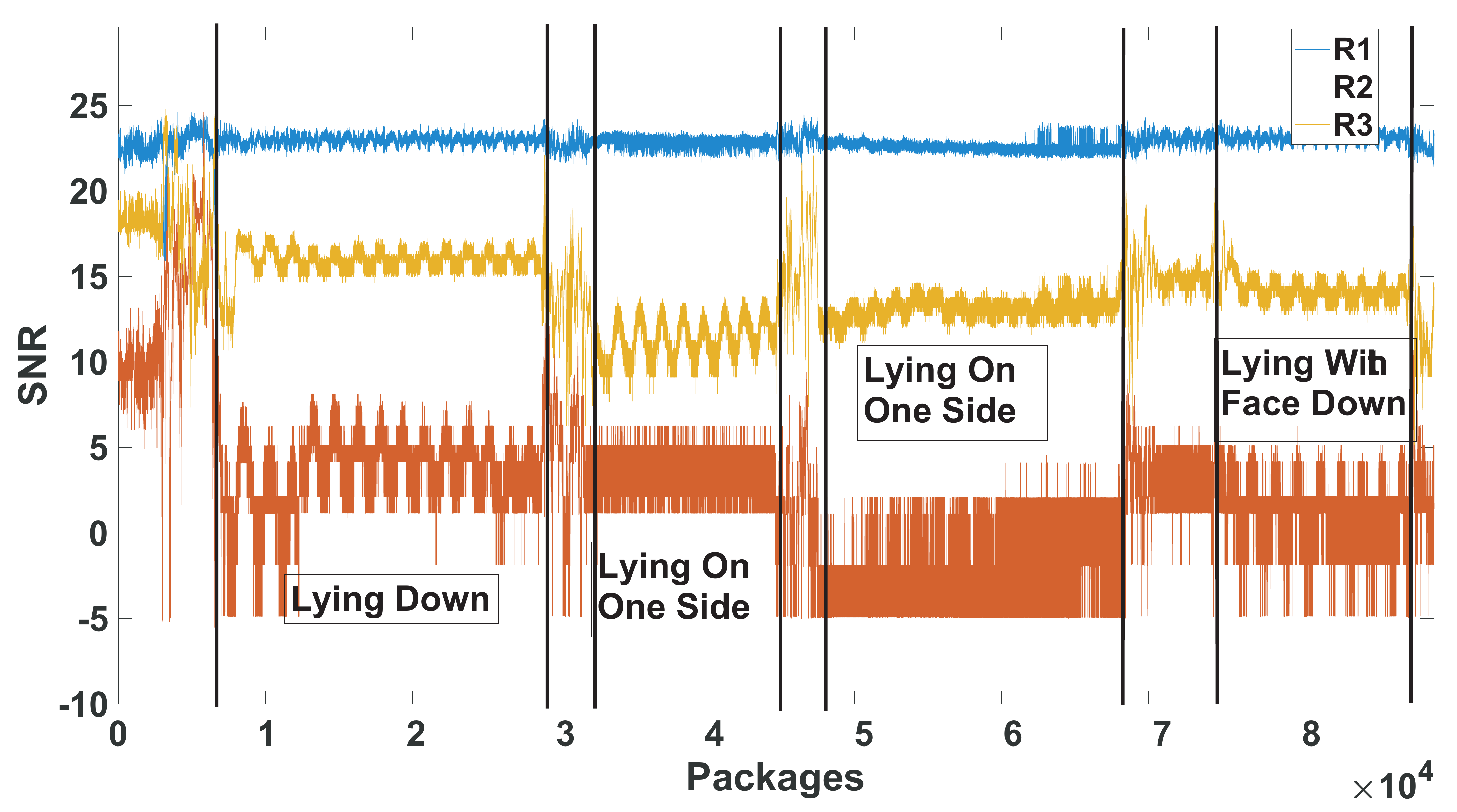}
		\end{minipage}
	}
	
	\quad
	\subfloat[]{\label{plan3}
		\begin{minipage}{0.5\linewidth}
			\centering
			\includegraphics[width=1\textwidth]{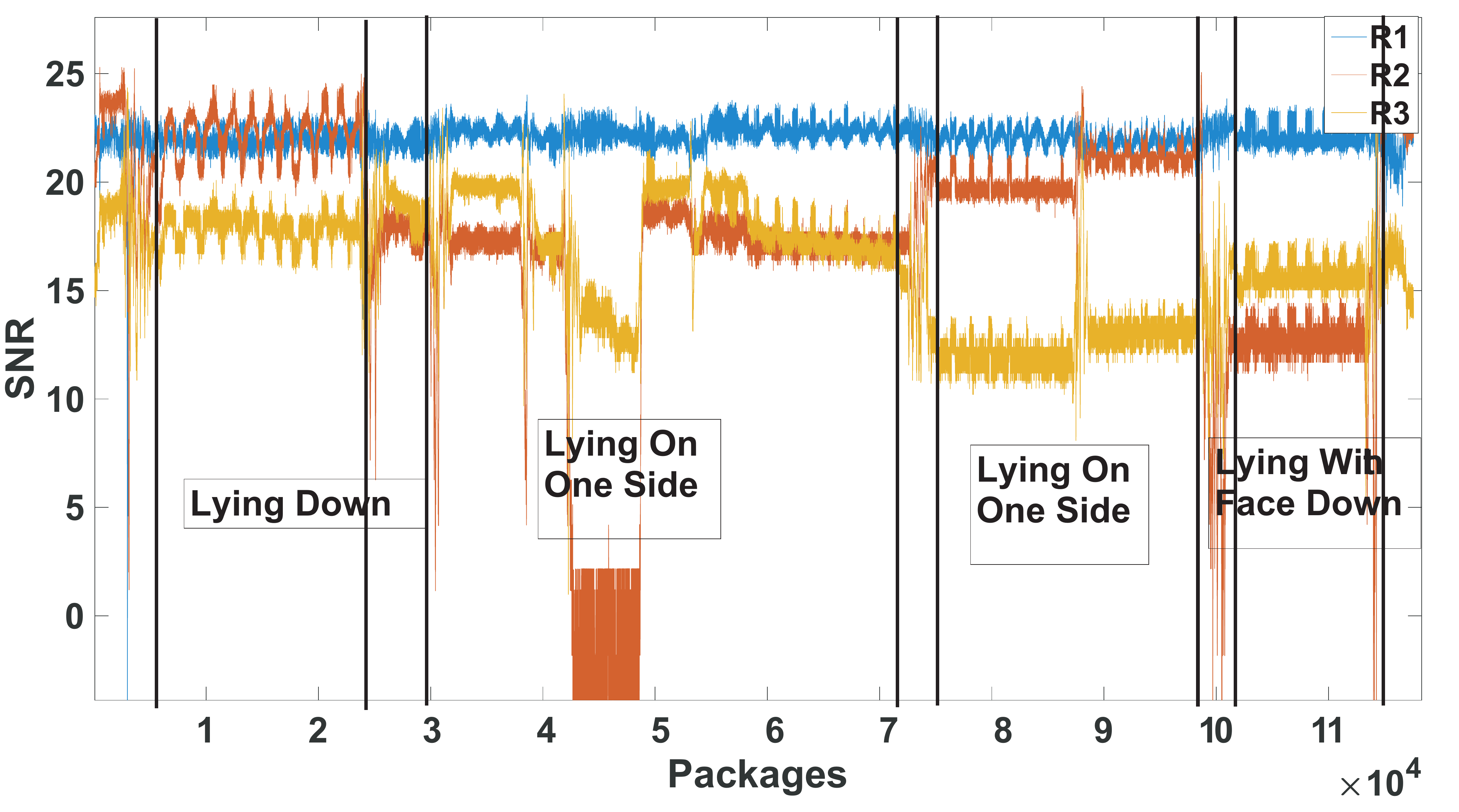}
		\end{minipage}
	}	
	\quad
	\caption{(a)Detection results of setting1; (b)Detection results of setting2; (c)Detection results of setting3.}
\end{figure*}

\textbf{Breath indeed affects channel response and such effect depends on the experimental settings:}

We confirm that breathing in terms of signal variations has been recorded by all settings, indicating that breath indeed affects channel response.
Different antennas settings bring different results. For setting 1, R3 (the yellow one) performs well for breath detection in all sleeping postures, and other antennas can also detect respiration. For setting 2 and 3, R2 (the red one) performs very well for lying down and lying with face down, but bad for other postures. R3 performs better with setting 2 than setting 3, in these two settings, the position difference of the transmitting antenna is not a lot.

Based on these, we can find that it is possible to detect the breathing in different sleeping postures by adjusting the antennas setting using only one antenna pair.

\textbf{The guidance provided by Fresnel theory is limited:}

Breath in the first Fresnel zone or on the LOS of antenna pair can indeed obtain good channel response with two sleeping postures (lying down and lying with face down), the same as described in previous work\cite{zhang2018fresnel}. However, it does not perform well in other sleeping postures (lying on one side), as shown in Fig.\ref{plan2} and \ref{plan3}, R2 (the red one). And hence we need to adjust the position of the abdomen to get a distinct respiratory waveform, different from described in related work\cite{zhang2018fresnel}. One possible reason is that the chest displacement during respiration in mediolateral dimensions is too small to detect.

In particular, for setting 1, when lying down or lying with face down, the direction of abdominal/thoracic motion caused by breathing is nearly parallel to T1-R3. Based on the Fresnel region theory, the effective displacement of this case is tiny, the performance of T1-R3 should not be excellent. However, T1-R3 is the best stream for all three streams, even if we adjust the height of T1-R1 to make the distance between T1-R1 equals to T1-R3, T1-R3 is still the best choice. This phenomenon is different from the description of the previous work \cite{wang2016human}, according to Wang.et.al's theory, T1-R1 should perform better than T1-R3 when human lying down, because of body orientation is 0 degree for T1-R1, but nearly 90 degrees for T1-R3.

The reason of this may relate to obstacles, since the antenna is placed on the shelf, there is a plastic plate blocking in the middle of T1-R3. The plastic plate blocks part of the direct signal, making the R3 more sensitive to motion and enhancing the performance. In the experiment, we also find that moving the position of one receiving antenna can affect the performance of the other two fixed receiving antennas.

In summary, Fresnel theory is beneficial, but it limited by other factors, we use our findings and Fresnel theory to enhance detection performance in section \ref{Sect:sys}.

\section{System Design}
\label{Sect:sys}
\begin{figure}
	\centering
	\includegraphics[width=1\columnwidth]{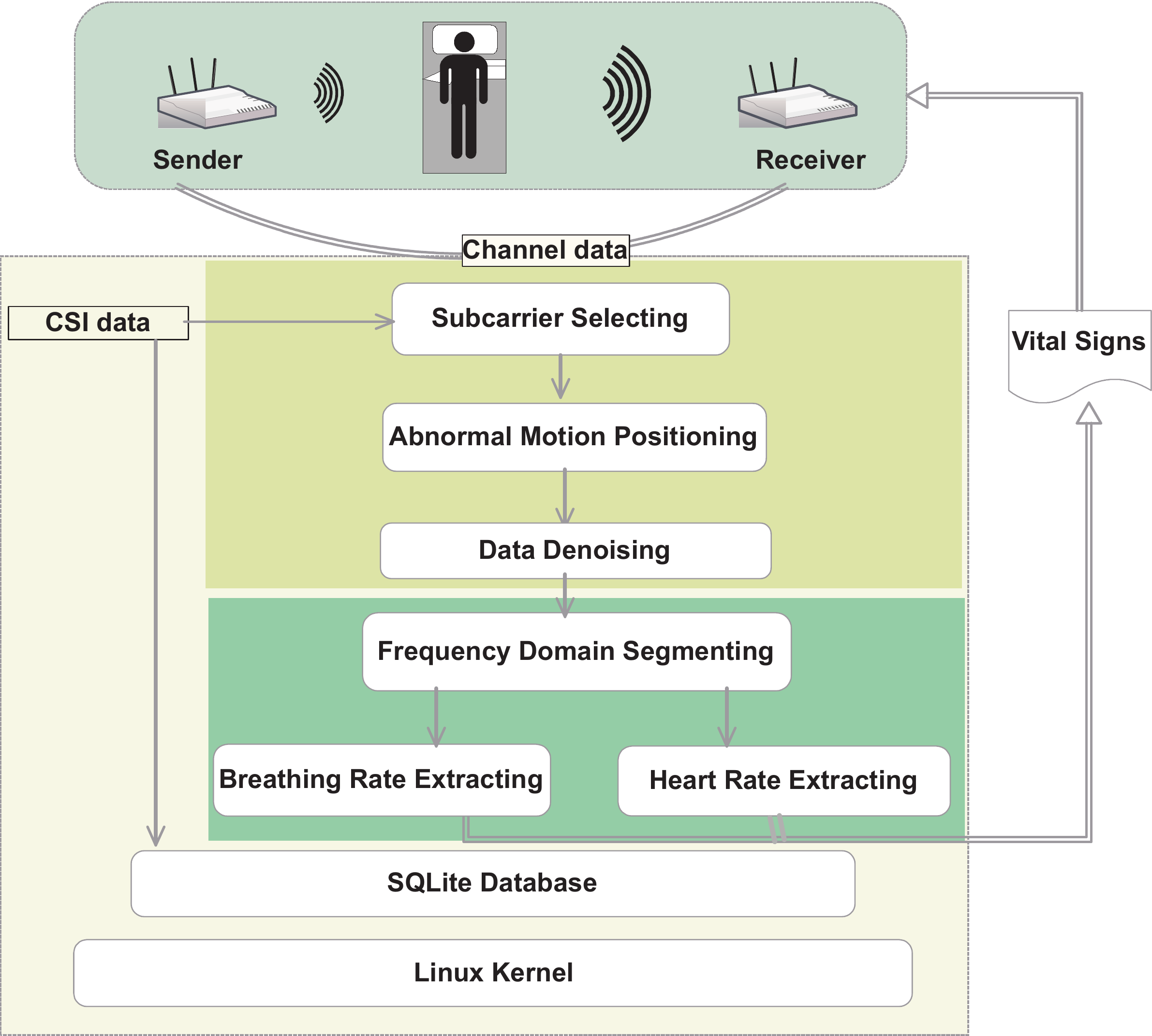}
	\caption{System architecture}
	\label{fig:system}
\end{figure}

In this section, we present the system design, as shown in Fig. \ref{fig:system}, we send the received CSI data to the real-time system for processing. We select the best performing subcarrier by subcarrier selecting mechanism. Then by denoising and frequency domain segmenting, we divide the data into two parts, mainly including breathing and mainly containing heartbeat, respectively. Finally, we extract the breathing rate and heart rate separately. We also store the raw CSI data in the database.

\subsection{Hardware Setting}
We choose setting1 to build our prototype system. We first calculate the Fresnel diffraction model of T1-R3, by adjusting the distance of T1-R3 and their position from the bed, to keep the bed in a sensitive surveillance area. We leave the bed outside the second Fresnel zone of T1-R3, but not too far away, and then place a lead sheet under T1 to enhance its sensitivity. Although our system only needs one data stream, we must place the other two antennas. This is due to that the CSI tool must receive the data stream of three receiving antennas at the same time to collect the CSI data.

Moreover, we find that no matter how the signal is blocked, there is always a data stream with an SNR around 20, which is the least sensitive. If we put three antennas together, there is one least sensitive data stream, and it's not fixed. Every time we restart the system, we need to find the best data stream. Many studies have not noticed this phenomenon; and hence three receiving antennas are placed together \cite{zhang2018fresnel}, which degrades the system performance.

\subsection{Data Processing}
\label{Sect:dat}
We use the CSI-tool to extract CSI data from received WiFi signal, and then send data to the monitor computer for processing.
The workflow is explained as follows;

{\bfseries \noindent Subcarrier Selecting.} Different subcarriers have different central frequencies, and for the same motion, they may have different performances. Therefore, it is essential to choose proper subcarriers that can better capture the breathing. According to previous experience\cite{gu2018your}, we choose the subcarrier with the most significant variance.

{\bfseries \noindent Data Denoising.} Received CSI data contains a lot of interference noise due to equipment and environmental factors. In the preprocessing module, we choose the Hampel filter to filter out the outliers which have significantly different values from other neighboring CSI measurements.

{\bfseries \noindent Frequency Domain Segmenting.} The CSI data with the frequency range related to people's normal heart rate range (i.e., 60bpm to 120bpm, which corresponds to 1Hz to 2Hz) will be sent for our heart rate estimation. And the CSI data with the frequency range related to people's normal breath rate range (i.e., 15bpm to 30bpm which corresponds to 0.25Hz to 0.5Hz)  will be used for our breath rate estimation. We use Butterworth bandpass filters to separate these two kinds of data.

\begin{figure}
	\centering
	\includegraphics[width=1\columnwidth]{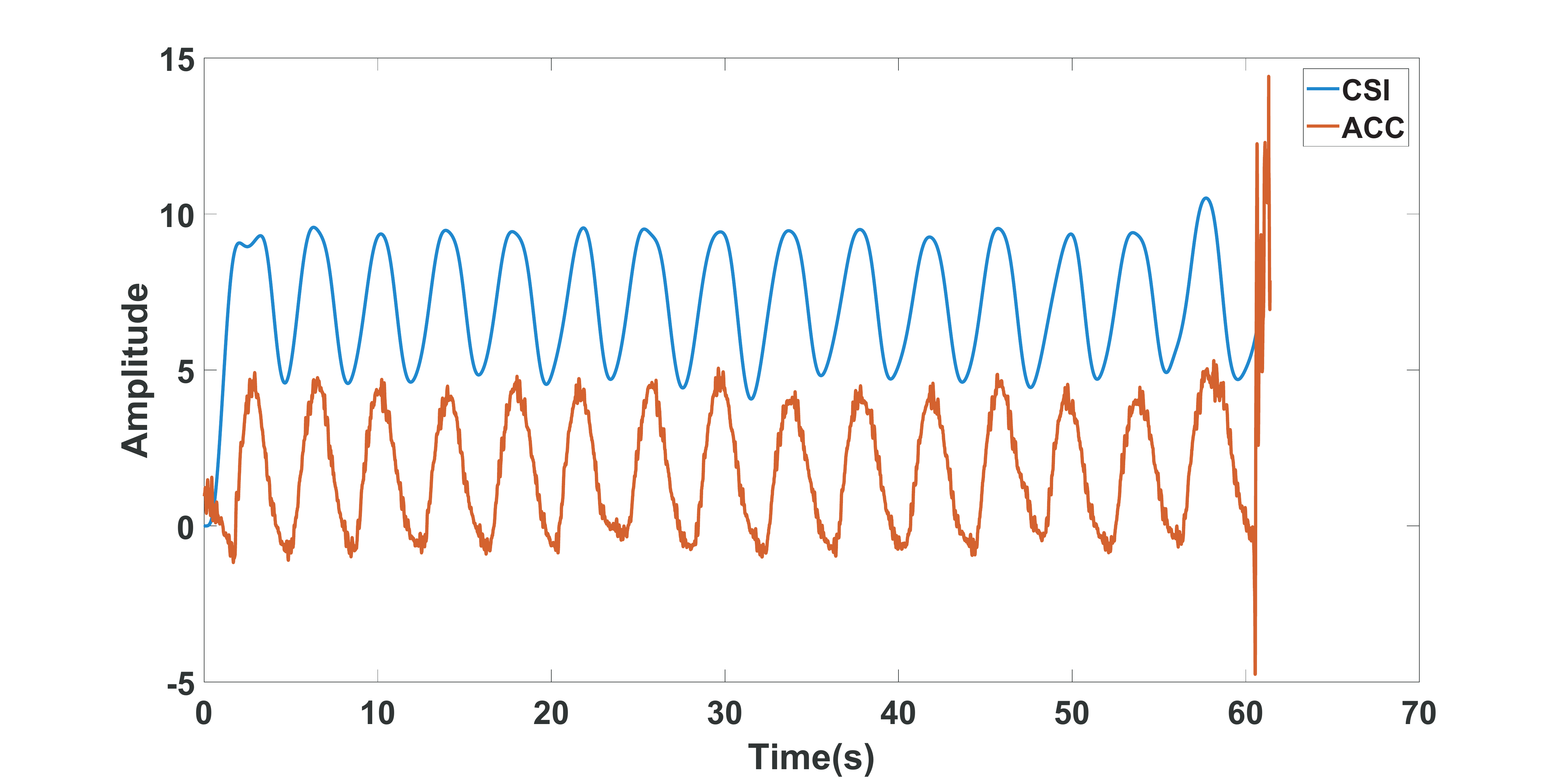}
	\caption{Comparison of processed CSI and ACC sensor readings for breathing.}
	\label{fig:breath}
\end{figure}

\begin{figure}
	\centering
	\includegraphics[width=1\columnwidth]{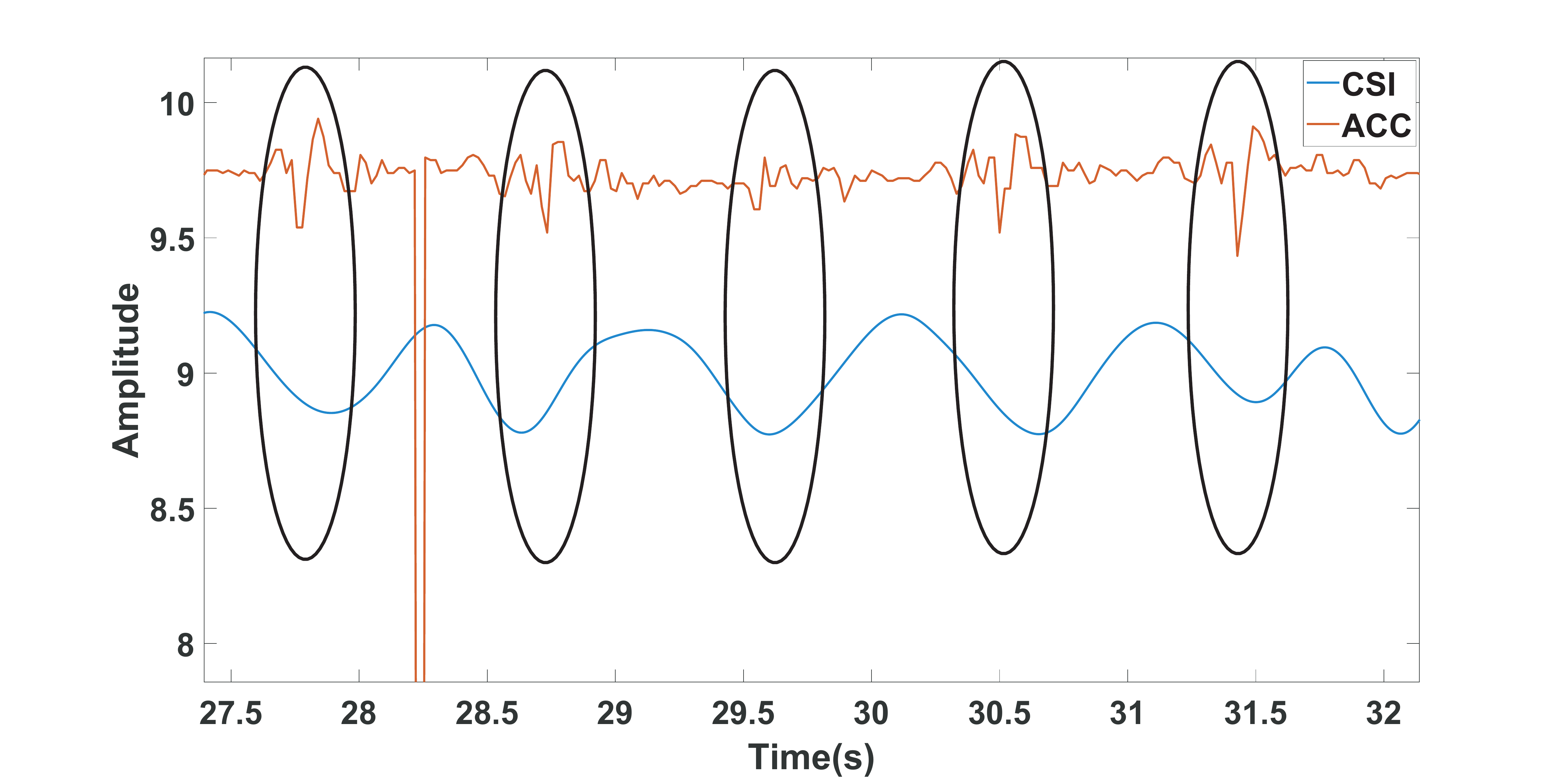}
	\caption{Comparison of processed CSI and ACC sensor readings for heart beat}
	\label{fig:heart}
\end{figure}

{\bfseries \noindent Vital Sign Extracting.} We use a bandpass filter to obtain data containing mainly heart rate and data providing primarily respiratory rate, and then extract heart rate and respiratory rate by FFT. We design a real-time system to process and display these vital signs in real-time use Matlab, and the system can manually adjust the FFT time threshold. If the accumulated data reaches the threshold time, the vital signs extraction section starts working, in general, 40s of data is enough to ensure accuracy. Considering the requirements of real-time systems and the weak multi-threading capabilities of Matlab, our extraction algorithm favors simple and efficient FFT.

The CSI data mainly containing the respiratory information obtained by the bandpass filter is compared with the acceleration sensor data placed on the abdomen as shown in the Fig.\ref{fig:breath}. It can be seen that the CSI waveform is highly consistent with the respiratory waveform obtained by the acceleration sensor. Fig.\ref{fig:heart} compares the normalized CSI patterns to an acceleration sensor’s readings placed on the chest, the occurrence of a heartbeat on the accelerometer is consistent with the detection result of CSI. It is indicating that the normalized CSI obtained from WiFi signals could be utilized to extract the fine-grained heart motions and breathing motions. We use Matlab to write a real-time system to monitor vital signs in real-time.

\section{Performance Evaluation}
\label{Sect:per}
\subsection{Hardware Setup}
\begin{figure}
	\centering
	\includegraphics[width=0.9\columnwidth]{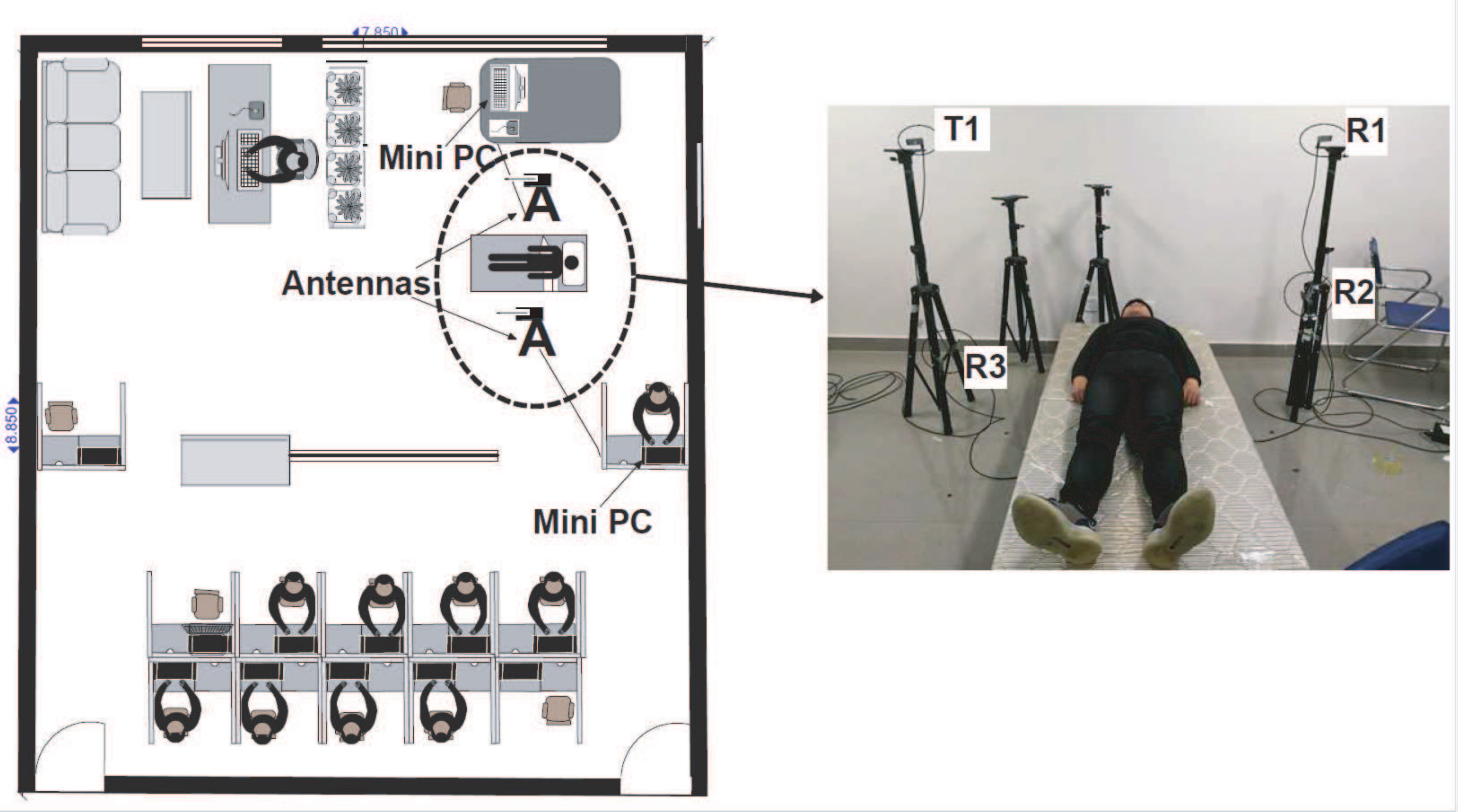}
	\caption{Illustration of the prototype system.}
	\label{fig:prototype}
\end{figure}

We use off-the-shelf hardware devices to implement the proposed system. Specifically, we use two miniPCs as the sending and the receiving devices, and their network cards are Intel Link 5300 WiFi NIC. The miniPCs have 2.16GHz Intel Celeron N2830 processor with 2GB RAM and Ubuntu OS in version12.04.

Our antenna settings are shown in Fig.\ref{fig:prototype}, the distance from T1 to R3 is 80cm, we place a lead sheet under T1 to enhance the performance of R3, and there is no blocking between R1 and T1. R1 has the highest SNR and the worst sensitivity in actual use. We only use R3 to monitor vital signs, the purpose of placing two other antennas is to ensure that the signal-to-noise ratio of R3 is not the highest. The sending speed of our transmitting equipment is 500 packets/second.

\subsection{Experimental Methodology}
We sout{conduct} experiment in the lab environment shown in Fig.\ref{fig:prototype}, a total of five participants join in the experiment (three male and two female students) whose ages rank from 21 to 26. These five participants are university students who volunteer for the experiments. In the experiments, we do not limit the normal activities of other peoples in the room.

Each participant performs an actual test of 30 minutes in different sleeping postures (prone, supine, facing left recumbent and facing right recumbent). Different from the previous work\cite{liu2018monitoring}, we do not use the metronome to control the participant's respiratory rate, and we don't need to use a directional antenna to detect heart rate under LOS conditions.

We use a real-time system to read our measured heart rate and respiratory rate directlysout. And the ground truths of breathing and heart rates are monitored by the accelerometer placed on the abdomen and a fingertip pulse oximeter, respectively. We use the difference between the measured value of the system and the measured value of the accelerometer as an error to characterize the performance.

\subsection{Evaluation Result}

\begin{figure}
	\centering
	\includegraphics[width=1\columnwidth]{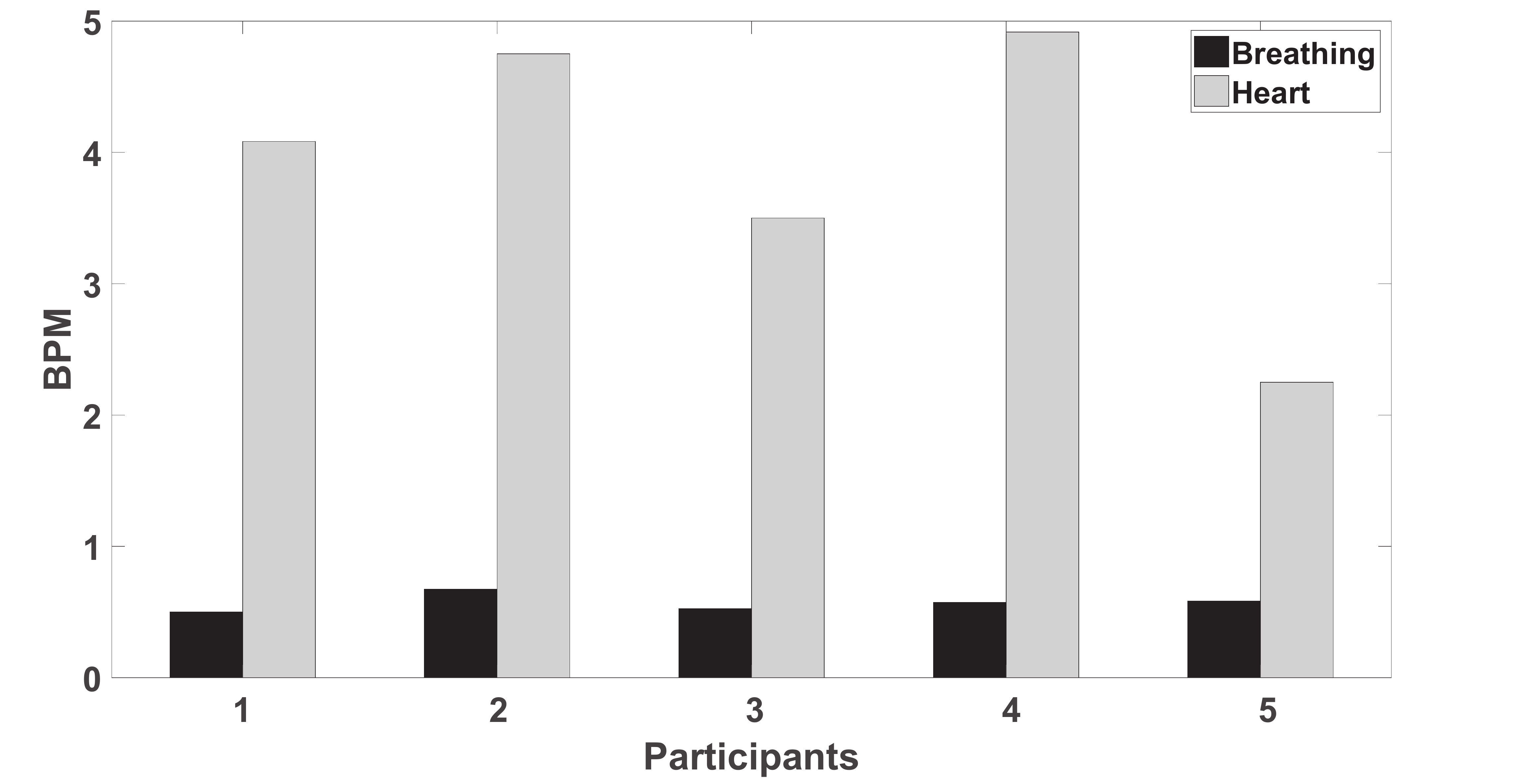}
	\caption{Illustration of the vital sign (breathing and heart rate) monitoring error of different participants.}
	\label{fig:participants}
\end{figure}

\begin{figure}
		\centering
		\includegraphics[width=1\columnwidth]{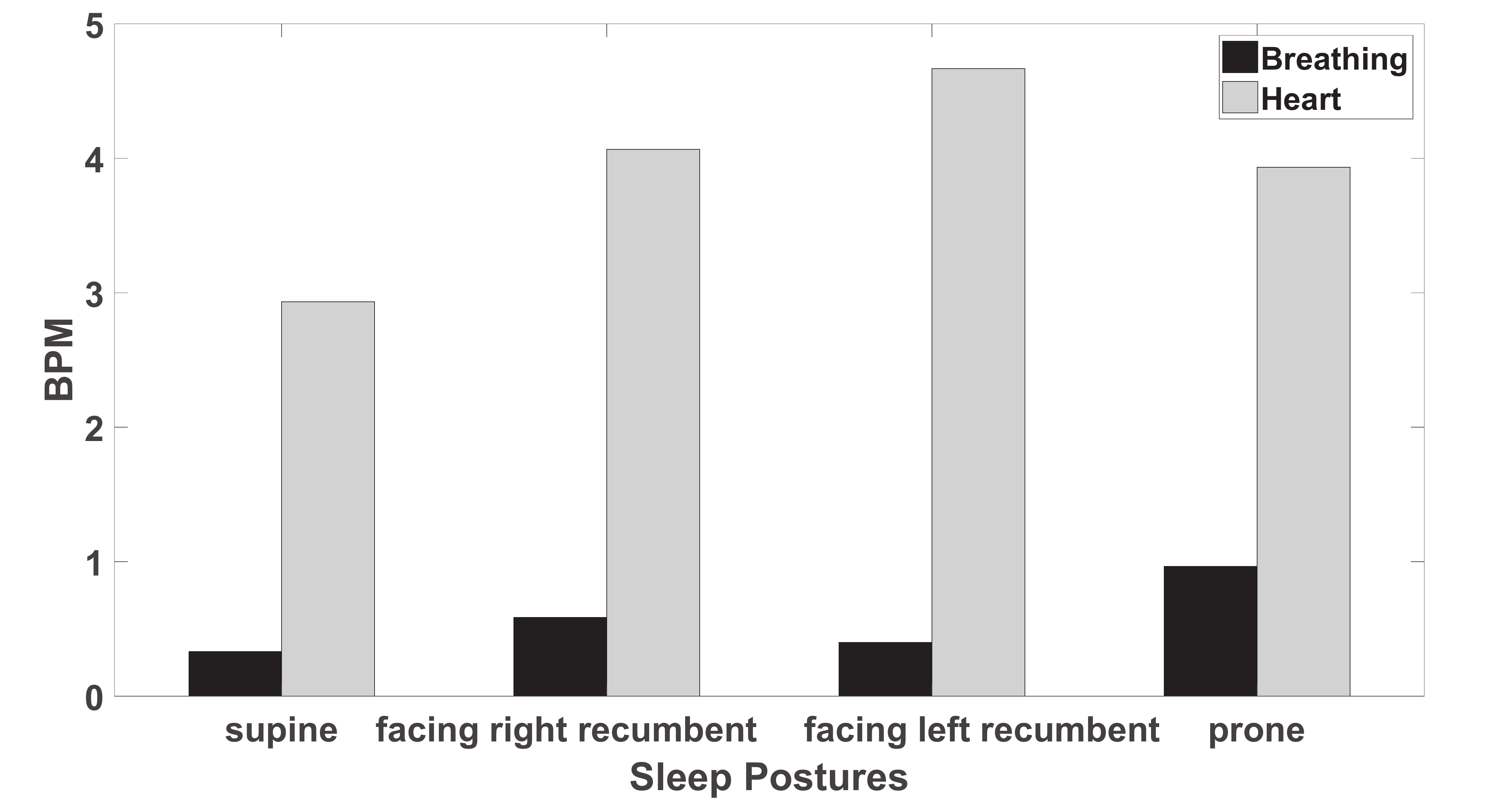}
		\caption{Performances of different sleep postures.}
		\label{fig:postures}
\end{figure}

We evaluate the overall performance of breathing and heart rate estimation under different sleep posturessout. The final result is an average error of 0.575 bpm (beat per minute) for detecting breath, 3.9 bpm for detecting heart rate. The overall accuracy is 96.636\% and 94.215\% ,respectively.

Fig.\ref{fig:participants} illustrates the vital sign (breathing and heart rate) monitoring error of different participants; they have different body types, which result in different final results. However, in general, our system has high accuracy in detecting respiration, and the error in detecting heart rate is also within the acceptable range in a non-clinical environment.

Fig.\ref{fig:postures} illustrates the vital sign (breathing and heart rate) monitoring error of different sleep postures. From the results, we can observe that the monitoring error in supine posture is relatively small. The facing left recumbent posture is with the largest error in monitoring heart rate, and the pron posture is with the most significant monitoring error in terms of breathing rate. In general, our system can accurately monitor vital signs with different sleeping postures.

\section{Conclusion And Future Work}
\label{Sect:con}
In this paper, we showed that we could use WiFi signals to track people's vital signs (breathing and heart rate) for different sleep postures using only one pair of WiFi devices. In particular, our system exploited CSI extracted from WiFi physical layer to detect the minor motions caused by breathing and heartbeat.
We set up the antennas based on Fresnel diffraction model and signal propagation theory, which enhanced the detection of weak breathing/heartbeat motion. We implemented a prototype system using the off-shelf devices and a real-time processing system to monitor vital signs in real-time.  The experimental results indicated the accurate breathing rate and heart rate detection performance.
In the future, we plan to design an abnormal motion positioning algorithm based on regularity detection, which can accurately position the range of abnormal motions (turn over and get up etc.) and classify the abnormal motions to provide more detailed sleep information.

\bibliographystyle{IEEEtran}
\bibliography{Vital}
\end{document}